\documentclass[notitlepage,nofootinbib,amsmath,longbibliography,amssymb,11pt,groupedaddress]{revtex4-1}
\usepackage{amsmath,amssymb,graphicx}
\usepackage{epsf}
\usepackage{epstopdf}
\usepackage {amssymb}
\usepackage{color}
\usepackage[colorlinks=true, linkcolor=blue, citecolor=magenta, hyperfootnotes=true]{hyperref}
\newcommand{\nc}{\newcommand}
\nc{\ba}{\begin{eqnarray}}
\nc{\ea}{\end{eqnarray}}
\newcommand\be{\begin{equation}}
\newcommand\ee{\end{equation}}

\newcommand{\calN}{{\cal{N}}}
\nc{\x}{{\bf{x}}}
\nc{\theo}{{\hat{\theta}}}
\nc{\dat}{{\hat \mu }}

\begin{document}
\nocite{TitlesOn}
\title{Bayesian forecasting with information theory}
\author{Mohammad Hossein Namjoo}
\email{mh.namjoo@ipm.ir}
\affiliation{School of Astronomy, Institute for Research in Fundamental Sciences (IPM),\\ Tehran, Iran, P.O. Box 19395-5531}

%\date{\today}

\begin{abstract}
Forecasting techniques for assessing the power of future experiments to discriminate between theories or discover new laws of nature are of great interest in many areas of science. In this paper, we introduce a Bayesian forecasting method using information theory. We argue that mutual information is a suitable quantity to study in this context. Besides being Bayesian, this proposal has the advantage of not relying on the choice of fiducial parameters, describing the ``true" theory (which is {\it a priori} unknown), and is applicable to any probability distribution. We demonstrate that the proposed method can be used for parameter estimation and model selection, both of which are of interest concerning future experiments. We argue that mutual information has plausible interpretation in both situations. In addition, we state a number of propositions that offer information-theoretic meaning to some of the  Bayesian practices such as performing multiple experiments, combining different datasets, and marginalization.

\end{abstract}

\maketitle
\tableofcontents

\section{Introduction}
\label{Sec:intro}

Forecasting the capability of a future experiment to distinguish different theories is of interest in many areas of science. This assessment allows us to judge about the cost and benefits of an experiment before it is conducted. It also provides a way to estimate what specifications are needed for an experiment if one is aimed at achieving certain constrains on the parameter space. A number of forecasting methods --- mostly within the frequentist framework --- exist, the prominent one being the Fisher analysis \cite{fisher1922mathematical}. Despite its simplicity and its wide range of applicability, the Fisher analysis has limitations, as it relies on the Gaussianity assumption of the probability distributions. Additionally, this method, along with many others, requires the selection of fiducial parameters to describe the ``true” theory, which is {\it a priori} unknown.

Recent technical developments made it feasible to consider the Bayesian method of inference (as opposed to the frequentist method) as a conceptually clear way to assign probabilities to the theories of interest and directly compare them once the outcome of an experiment becomes available. As a result, the Bayesian paradigm has recently attracted much attention, especially in physics. However, only a few forecasting methods are proposed within this framework. Our objective in  this paper is to introduce a Bayesian forecasting method using the concepts borrowed from information theory. In particular, we argue that {\it mutual information} is a reasonable quantity to study for Bayesian forecasting, due to its clear meaning and its desirable properties. We argue that mutual information can be used in both situations of interest in science, namely, for parameter estimation and model comparison. Besides being Bayesian, this proposal --- in contrast with  many other forecasting methods --- has the advantage that it does not rely on the assumption of picking out a specific theory (i.e., the choice of fiducial theory) nor does assume any particular form of probability density.

The rest of the paper is organized as follows. In Secs.~\ref{Sec:Bayes} and \ref{Sec:Info} we briefly review the Bayesian probabilistic inference and the information-theoretic framework, respectively, and introduce our notation. Sec.~\ref{Sec:Info-bayes} is devoted to the main task of this paper, i.e., justifying the application of the information-theoretic framework for Bayesian forecasting. In Sec.~\ref{Sec:propositions} we state a few propositions that may be viewed as the information-theoretic interpretation of some  Bayesian exercises. We conclude along with some further remarks in Sec.~\ref{Sec:conclusion}.

\section{Bayesian inference}
\label{Sec:Bayes}

We start with a brief overview of Bayesian inference. A more detailed discussion may be found in the standard textbooks such as \cite{bolstad2016,gelman2013bayesian} (see also Refs.~\cite{Trotta:2008qt,Verde:2009tu}). We denote the random variable that is going to be measured by $\dat$ and a specific realization of it by $\mu$. The theories are parametrized by another random variable that we denote it by $\theo$ and its realization by $\theta$.\footnote{We loosely use the same notation to denote the parameter and the theory that is described by the parameter.} Concerning parameter estimation, when $\mu$ is the outcome of an experiment, the ultimate goal in the Bayesian framework is to estimate the posterior probability density $p(\theta|\mu)$ via the Bayes theorem
\ba 
\label{Bayes}
p(\theta|\mu) = \dfrac{p(\mu|\theta) p(\theta)}{p(\mu)}\, ,
\ea 
where $p(\mu|\theta)$ is the likelihood density (i.e., the probability density of the measured value $\mu$ given the theory $\theta$) and $p(\theta)$ is the prior probability density. $p(\mu)$ is the normalization factor and is given by 
\ba 
p(\mu) = \int p(\mu|\theta) p(\theta) d\theta.
\ea 
To simplify the notation, we use a univariate-like symbols but the entire analysis in this paper is also applicable to the multivariate case. 

When  model selection, rather than parameter estimation, is concerned, a discrete version of the  Bayes formula applies. Denoting the models under consideration by $M_i$ for $i=1,...,N$ (where $N$ is the total number of models) we have
\ba 
P(M_i|\mu) = \dfrac{p(\mu|M_i)P(M_i)}{\sum_{j=1}^N p(\mu|M_j)P(M_j)}\ ,
\ea 
Throughout this paper, we denote probabilities and probability densities by $P$ and $p$ (upper and lower case), respectively. 

\section{Information theory framework}
\label{Sec:Info}
We now turn to the relevant definitions in information theory. We state some standard properties of different variables, the proofs of which may be found in related textbooks (see e.g., \cite{book1,book2}). We follow the notation of Sec.~\ref{Sec:Bayes}, hence e.g., if the realization of a random variable is denoted by $x$, the corresponding random variable will be denoted by $\hat x$. 

 In the situations where discrete outcomes are expected, one can define {\it entropy}, which quantifies the uncertainty about the outcome,  by
\ba 
\label{entropy}
H \equiv -\sum_i P_i \ln P_i \, ,
\ea 
where $P_i$ is the probability of realization of the $i$-th possible outcome. 
A natural generalization to the case with a continuous outcome is the {\it differential entropy} defined by
\ba 
\label{diff-entropy}
H(\hat x) \equiv  - \int p(x) \ln p(x) \, dx.
\ea  
Note, however, that this generalization does not lead to a quantity that has all desirable properties of an entropy. In particular, it can be negative. Nonetheless, we can still interpret it as information, as long as differences between differential entropies are considered. Note that the probability distribution in Eq.~\eqref{diff-entropy} can be a conditional probability density. In this case, it measures the uncertainty in one variable given a specific realization of another random variable:
\ba 
\label{diff-cond-realization}
H(\hat x|y) \equiv  - \int p(x|y) \ln p(x|y) \, dx.
\ea 
 It would then be useful to  define conditional entropy; the expected entropy of a random variable when another random variable is given
\ba 
\label{cond-entropy}
H(\hat x|\hat y) \equiv \int H(\hat x|y) p(y) \, dy = -\int p(x|y) p(y) \ln p(x|y) \, d x d y.
\ea 
A more interesting quantity is {\it mutual information (MI)} which is defined by 
\ba 
I(\hat x, \hat y) \equiv  \int p(x,y) \ln \left( \dfrac{p(x,y)}{p(x)p(y)} \right)\, dx dy .
\ea 
Intuitively,  MI measures the information that one expects to gain about one random variable when another is measured. 
MI is symmetric $\big(I(x,y)=I(y,x)\big)$ and non-negative. It is zero if and only if $x$ and $y$ are independent. 
Similar to the differential entropy, one can define the MI between two variables conditioned on the realization of a third variable by 
\ba 
\label{MI-cond-realization}
I(\hat x,\hat y| z)\equiv \int p(x, y| z)   \ln \left( \dfrac{p(x, y| z)}{p(x| z)p(y|z)} \right) \,  dx dy   .
\ea 
Moreover, the expected MI between two variables when a third random variable is given, may be quantified by {\it conditional mutual information} defined by
\ba 
\label{MI-cond}
I(\hat x,\hat y| \hat z) \equiv \int I(\hat x,\hat y| z)  p(z) \, dz .
\ea 

For the future reference, we also introduce the {\it relative entropy} (or Kullback-Leibler divergence) between two probability distributions $p(x)$ and $q(x)$ by \cite{Kullback:1951zyt} 
	\ba 
	\label{KL}
	D\big(p(x)||q(x)\big) = \int p(x) \ln \left( \dfrac{p(x)}{q(x)} \right) dx.
	\ea 
	Relative entropy measures the difference between two probability distributions.

In all of the above relations if one of the random variables is discrete, the corresponding integral is replaced with the summation over all possible discrete outcomes.

\section{Information theory approach to Bayesian forecasting}
\label{Sec:Info-bayes}

We are now prepared to propose a Bayesian forecasting method using  information theory. As stated earlier, MI is the  information one expects to gain about one variable when another variable is measured. This is in fact the concept that a Bayesian would like to engage with regarding a future experiment: For a given experiment and a hypothetical measurement of a random variable, we would like to ask to what degree the uncertainty in the space of possible theories will be reduced? In other words, we would like to ask how much information we expect to obtain about the theories from a measurement? The connection between this question and the interpretation of MI, naturally suggest to use MI for forecasting and it is indeed the central quantity that we work with in this paper. 
Note the Bayesian nature of the MI between the two random variables $\dat$ (the variable to be measured) and $\theo$ (the variable that labels theories): Using the relation between joint and conditional probabilities, it is clear that MI depends on the likelihood density as well as on the prior probability density (or, from the Bayes theorem Eq.~\eqref{Bayes}, on the posterior probability density):
\ba 
\label{MI-conditional}
I(\theo, \dat) =\int p(\mu|\theta) p(\theta) \ln \left( \dfrac{p(\mu|\theta)}{p(\mu)} \right) d\theta d\mu=
\int p(\theta | \mu) p(\mu) \ln \left( \dfrac{p(\theta| \mu)}{p(\theta)} \right)d\theta d\mu.
\ea

 Our proposal is to use MI as a criterion to decide whether a future experiment is able to discriminate different theories or not.  An immediate advantage is that MI does not depend on the specific choice of a fiducial theory.\footnote{ We refer to the theory described by the fiducial parameters as the {\it{fiducial theory}}. This is consistent with the Bayesian nomenclature while, in the frequentist framework, it may be called the {\it{fiducial model}}.} 
 	This is in contrast with  many other forecasting methods where a fiducial theory must be assumed (e.g., to estimate the Fisher information matrix or to generate mock data). 
Moreover, note that MI can be used for any normalizable probability distribution and, e.g.,  does not rely on the Gaussianity assumption (which is the basic limitation of the Fisher analysis).

To obtain further intuition about the meaning of MI and its relevance to Bayesian forecasting, consider a hypothetical measurement of a random variable $\dat$ with the outcome $\mu$. Using this data, the prior probability density $p(\theta)$ may be updated to a posterior probability density $p(\theta|\mu)$. The reduction of the uncertainty in the space of possible theories (or the information gained about possible theories)  may be quantified by the difference between prior and posterior entropies
\ba 
\Delta H (\theo|\mu) \equiv  H(\theo) - H(\theo|\mu)\ ,
\ea 
where $H(\theo)$ and $H(\theo|\mu)$ are given by Eqs.~\eqref{diff-entropy} and \eqref{diff-cond-realization} using prior and posterior probability densities, respectively. If forecasting for a future measurement is concerned, one does not have access to the measured value $\mu$. Instead, one can obtain the expected value of $\Delta H (\theo|\mu)$ over all possible realizations of $\dat$. But this is identical to MI and we have 
\ba 
\label{MI_DeltaH}
I(\theo,\dat) = \int \Delta H (\theo|\mu)  \, p(\mu) d\mu = H(\theo) -H(\theo|\dat).
\ea 
where $H(\theo|\dat)$ is the conditional entropy defined in Eq.~\eqref{cond-entropy}. Therefore, we arrive at the following important conclusion:
\begin{quote}
	{\it Mutual information quantifies the expected reduction in uncertainty (or the expected gain of information) about possible theories resulting from the conduct of an experiment. }
	\label{MI-meaning}
\end{quote}

We have seen that MI naturally arises from a Bayesian view to a hypothetical experiment. However, the relative entropy between the prior and posterior probability densities has also been considered as an information-theoretic quantity for forecasting (see e.g., \cite{March:2011rv,Paykari:2012ne,Amara:2013swa,Seehars:2014ora,Zhao:2017cud}). Relative entropy is asymmetric in exchanging the two distributions (see Eq.~\eqref{KL}), making it hard to assign an intuitive interpretation to it. Nonetheless, it is stated that relative entropy measures the difference between the two distributions. However, since --- for the purpose of forecasting --- there is no actual measured value, the posterior probability density is not yet determined. This enforces one to study relative entropy assuming a specific realization of the data, which is a downside of using this quantity. One can improve the situation by averaging over the measured value, similar to what we did above for $\Delta H(\hat \theta | \mu)$ (which, according to Eq.~\eqref{MI_DeltaH}, led to MI). Interestingly, MI emerges once again by this procedure. That is, we have 
\ba 
I(\theo,\dat) = \int D \big(p(\theta|\mu)||p(\theta) \big) \, p(\mu) d\mu. 
\ea

In the rest of this section, we provide additional intuition about MI and argue that it has further plausible interpretation in both situations of interest: parameter estimation and model selection. 

\subsection{Mutual information and parameter estimation}
When parameter estimation is of interest, one way to assess the significance of a future experiment from a Bayesian perspective is to determine how much the posterior probability density will spread in the parameter space around the “true” theory, compared to the spread of the prior probability density. We assert that MI captures this idea (albeit, in a way that is free from the choice of a ``true" theory). This is, indeed, a consequence of our statement in Sec.~\ref{MI-meaning} (about the meaning of MI) applied to the specific situation of parameter estimation: (1) MI measures the reduction in the uncertainty about theories; (2) uncertainty and dispersion of the probability distribution are closely related concepts. Thus, we contend, MI measures the reduction in the dispersion of the probability distribution of theories resulting from conducting an experiment.  
This aligns with the usage of differential entropy as a measure of dispersion that has been suggested e.g., in \cite{Erlander1980,theil1967}. Taking this suggestion for granted,  Eq.~\eqref{MI_DeltaH} then implies that MI is the difference between the prior and posterior dispersion.
 Therefore, we conclude that 
\begin{quote}
	\label{MI_param_estim}
	\it Concerning parameter estimation, the quantity $I(\theo,\dat)$ estimates the expected decrease in the dispersion of the posterior probability density compared to that of the prior probability density.
\end{quote}

A simple example illustrates the relation between MI and a more familiar measure of dispersion, i.e., the standard deviation. Denote the multivariate normal distribution for a vector of random variables $x$ with mean $\bar x$ and covariance matrix $\Sigma$ by $\calN_x (\bar x, \Sigma)$.
Consider a situation where the likelihood probability density may be estimated by a multivariate Gaussian distribution with  (unknown) mean $\theta$ (which also labels different theories) and  (known) covariance matrix $\Sigma_L$, $p(\mu|\theta) = \calN_\mu  (\theta,\Sigma_L)$. Further consider a Gaussian prior with mean $\theta_0$ and covariance matrix $\Sigma_{\text{pr}}$, $p(\theta) =\calN_\theta (\theta_0,\Sigma_{\text{pr}})$. Then, it is easy to show that MI is given by 
\ba 
\label{MI-posterior-width}
I(\theo,\dat) = -\dfrac{1}{2} \ln \dfrac{\det \Sigma_{\text{post}}}{\det \Sigma_{\text{pr}}}\, ,
\ea 
where $\Sigma_{\text{post}}$ is the posterior covariance matrix and is given by 
\ba 
\Sigma_{\text{post}}^{-1} = \Sigma_L^{-1} + \Sigma_{\text{pr}}^{-1}.
\ea 
Defining the {\it generalized standard deviation} as the square root of the determinant of the covariance matrix, from Eq.~\eqref{MI-posterior-width}, we conclude that $e^{-I(\theo,\dat)}$ is precisely the ratio of the generalized standard deviation of the posterior probability density to that of the prior probability density. This relation is  shown here only for the simple example of  purely Gaussian probability distributions. However, in many other examples we examined (but do not present in this paper), it is either exactly or approximately true.\footnote{Another example for which this statement holds true will be given in Sec.~\ref{Sec:propositions}. Also, for further evidence for the generality of this claim, see Ref.~\cite{JWST} that studies an involved, but realistic, example relevant to cosmology.} This suggests that this relation is much more generic and is expected to be true, at least approximately, in many situations of practical interest. In any case, this relation --- regardless of its  level of applicability --- is very helpful for deciding which range of MI values makes the conduction of an experiment valuable, much like the frequentist practice of converting $p$-values to the level of significance, using a Gaussian probability distribution. See the discussion of Sec.~\ref{Sec:conclusion}. 

 Note that $0<e^{-I}\leq 1$. The upper bound  may be interpreted as indicating that new data cannot increase uncertainties in the parameter space (i.e., the dispersion of the probability distribution always decreases as a result of update). In other words, by an experiment, information cannot be lost; it only can be gained. 

We conclude that MI is a plausible quantity to study for the purpose of forecasting concerning parameter estimation. In the next section, we study the interpretation of the same proposal applied to the case of model selection.

 We end this section by pointing out that the approach advocated here --- when is limited to the case of parameter estimation --- resembles what appear in recent engineering and computer science literature in the context of {\it parameter identifiability}. For example, see Refs.~\cite{Ebrahimian, PANT201566, capellari2017parameter}. We are expanding on these findings, introducing the concept to the physics community, and  providing additional insights that are particularly relevant to forecasting (including both parameter estimation and model selection situations).

\subsection{Mutual information and model selection}
We now turn our attention to model selection which is another situation of interest in Bayesian framework.\footnote{From the Bayesian perspective, model selection and parameter estimation are not conceptually different. The only distinction is that for model selection, a discrete variable is considered instead of a continuous one. However, this difference leads to assigning different meanings to MI in these two scenarios.} We will see that another intuitive interpretation of MI emerges in this case. 
Denote by $\hat M$ the random variable that the models of interest $M_i$ are its realizations.
 The likelihood density for each model may be determined by marginalization over all its parameters. That is
\ba 
\label{model_marginalize}
p(\mu|M_i) = \int p(\mu|\theta, M_i) P(\theta|M_i) d\theta.
\ea 
Prior entropy and conditional (posterior) entropy are given by (see Eqs.~\eqref{entropy} and \eqref{cond-entropy})
\ba 
H(\hat M)= -\sum_i P(M_i) \ln P(M_i)\, , \quad  H(\hat M|\dat) = -\int \sum_i P(M_i|\mu) p(\mu) \ln P(M_i|\mu) d\mu \ ,
\ea 
where $P(M_i)$ is the prior probability. 
MI is also given by
\ba 
\label{MI_model_selection}
I(\dat,\hat M)=\int \sum_i p(\mu|M_i) P(M_i) \ln \left(\dfrac{p(\mu|M_i) }{p(\mu)}\right) d\mu = H(\hat M) - H(\hat M|\dat).
\ea 
Since $\hat M$ is a discrete random variable, entropy, conditional entropy, and mutual information are all non-negative. It follows that $0 \leq I(\dat,\hat M) \leq H(\hat M)$.  $H(\hat M)$ can be interpreted as the total information that is in principle accessible (i.e., the prior uncertainty about the models under consideration). On the other hand, recall that $I(\dat,\hat M)$ measures the expected information accessible through the experiment. This allows us to give an intuitive meaning to the ratio $ I(\dat,\hat M)/H(\hat M)=1-H(\hat M|\dat)/H(\hat M)$ as follows:
\begin{quote}
	\label{MI_model_selec}
	\it Concerning model comparison, the ratio $I(\dat,\hat M)/H(\hat M)$ quantifies the expected fraction of information about the models that can be gained  from a given experiment. 
\end{quote}

We thus have seen that MI has a clear meaning generally (Sec.~\ref{MI-meaning}) and has plausible interpretation in both situations of model selection (Sec.~\ref{MI_model_selec}) and parameter estimation (Sec.~\ref{MI_param_estim}). 
\\\\
We end this section by comparing our approach with the proposals of  Refs.~\cite{Trotta:2007hy,Gessey-Jones:2023vuy} as a few other methods appeared in the literature for Bayesian forecasting, for the specific case of model selection (and not for  parameter estimation). A thorough study of advantages and disadvantages of each proposal is worth exploring but is beyond the scope of this paper. We only make a few comments that may suggest that the usage of MI is more beneficial. 

In Ref.~\cite{Trotta:2007hy} the  Bayes factor is utilized as the quantity that naturally emerges in the Bayesian framework. However, this proposal has the drawback that it relies on the choice of fiducial theory.
%\footnote{\red{The same drawback exists for the other standard method of analysis: generating mock data (which needs to ) and treats it like the real data, which is another straightforward method of Bayesian forecasting.} } 
This shortcoming is overcome in Ref.~\cite{Gessey-Jones:2023vuy} by marginalizing over the fiducial values and estimating the expected probability that the Bayes factor passes a certain threshold. We find it a reasonable solution to the aforementioned problem. 
However, the usage of MI seem to have the following two advantages, compared to the proposal of Ref.~\cite{Gessey-Jones:2023vuy}: First,  the analysis of Ref.~\cite{Gessey-Jones:2023vuy} is restricted to model selection whereas the usage of MI is suitable for both  model selection and  parameter estimation. This allows one to answer both kinds of questions within the same framework. Second, the computation of the expected probability that the Bayes factor is above a certain threshold, as suggested by  Ref.~\cite{Gessey-Jones:2023vuy}, needs to be repeated if one decides to change the threshold (or, more generally, decides to use a different criterion). In contrast, MI is estimated once. The threshold one requires for the adequacy of the experiment can be decided after the computation of MI.

\section{Information-theoretic interpretation of Bayesian practices}
\label{Sec:propositions}

In this section, we present some propositions that may offer information-theoretic insights into various circumstances commonly encountered in Bayesian inference. The proofs of propositions are simple and mostly follow from the definitions and the standard relations between conditional and joint probabilities. Each proposition is roughly stated to capture the corresponding intuition but is then made precise by a mathematical relation.

The first proposition makes precise how we learn about our theories when multiple experiments are taken into account. This is a very common situation in essentially all areas of  science. If multiple experiments are performed (or one experiment is performed multiple times) the information gained about possible theories generically differs from the sum of information gathered from individual experiments. This is due to the two effects. First, the cross-correlation between multiple datasets, as predicted by theories, may provide additional information that is not accessible if individual experiments are considered. Second, learning about theories from an experiment would be harder if it has been already  learned much about theories from previous experiments. This is because some information about theories is shared between different experiments. Clearly, these two effects act in opposite directions. In fact, restricting the discussion to two experiments, we have
\begin{quotation}
	{\bf Proposition 1.} (multiple  experiments):
	\\  Consider a set of theories that need to be tested through two experiments. The data from each experiment may contain redundant information, but there is also potentially  new information from cross-correlations between the outcome of different experiments, if it is predicted by the theories.
	 In equations, we have
	\ba 
	\label{disjoint-data}
	I(\{\dat_1,\dat_2\},\theo)= I(\dat_1,\theo)+I(\dat_2,\theo)+I(\dat_1,\dat_2|\theo)-I(\dat_1,\dat_2)\, ,
	\ea 
		where $ \dat_1 $ and $\dat_2$ are the two random variables that are going to be measured and $\theo$ parametrizes the theories under consideration.
\end{quotation}
%Note that the two datasets being drawn independently of each other does not imply that their corresponding random variables are independent. This is, while $\dat_1$ and $\dat_2$ are not {\it a priori} independent, (i.e., one has $p(\mu_1,\mu_2|\theta)=p(\mu_1|\theta)\, p(\mu_2|\theta)$) they correlate once the measurement is performed since they are both correlated with the random variable $\Theta$ (i.e., $p(\mu_1,\mu_2) =\int p(\mu_1,\mu_2|\theta) p(\theta) d\theta \neq p(\mu_1)\, p(\mu_2)$).

In Eq.~\eqref{disjoint-data}, the third term quantifies the information accessible through the correlation between the experiments (according to the theories). In contrast, the last term quantifies the information that is shared between the two experiments (which can be learned only once). In other words, the last term of Eq.~\eqref{disjoint-data}  describes the fact that it is harder to learn about theories if a lot is already learned.
%This proposition may be useful when  an experiment is expected to be repeated more than once or more than one experiment needs to be taken into account.  The former situation is clearly common in most areas of physics while the latter is typical in cosmology. In situations where MI from multiple datasets is too complex to estimate, the sum of  MI from individual datasets provides an upper bound on the total MI. Moreover, in situations where $\dat_1$ and $\dat_2$ are sensitive to different subsets of parameters $\theta$, the last term of Eq.~\eqref{disjoint-data} vanishes and MI reduces to the sum of MI from individual datasets.
 As a simple example that illustrates this intuition, consider a situation where $n$ independent samples are drawn from a normal distribution with a known standard deviation $\sigma_L$, and the goal is to estimate the mean of the distribution,  which we denote by $\theta$. That is, the likelihood density is given by $p(\mu|\theta)=\calN_\mu(\theta,\sigma_L^2)$. Further, assume a normal prior probability density with zero mean and standard deviation $\sigma_{\text{pr}}$, $p(\theta) = \calN_\theta (0,\sigma_{\text{pr}}^2)$. This situation is equivalent to measuring $n$ independent random variables $\{\dat_1,...,\dat_n\}$ only once, assuming the same likelihood densities $p(\mu_i|\theta)=\calN_{\mu_i}(\theta,\sigma_L^2)$, and the purpose is to estimate the true value of $\theta$. The total likelihood density in this scenario is given by $p(\{\mu_1,...,\mu_n\},\theta) = \prod_i \calN_{\mu_i}(\theta,\sigma_L^2)$. It is then easy to compute MI for this situation which given by
\ba 
\label{MI-n-measurements}
I(\{\dat_1,...,\dat_n \},\theo) = \dfrac{1}{2} \ln \big(1+n\dfrac{\sigma_{\text{pr}}^2}{\sigma_L^2}\big).
\ea 
For $n=1$, Eq.~\eqref{MI-n-measurements} is consistent with Eq.~\eqref{MI-posterior-width}. For $n>1$, this result is always less than $n I(\dat_1,\theta)$; the information gain grows only logarithmically (rather than linearly) for large $n$. This is due to the non-vanishing MI between different measurements (i.e., due to the last term of Eq.~\eqref{disjoint-data}):\footnote{Note that the third term of Eq.~\eqref{disjoint-data}, i.e., $I(\dat_1,\dat_2|\theo)$ vanishes in this example.} a new measurement is more predictable (and hence contains less information) when a lot is already learned by previous measurements. In fact, defining the {\it rate of information gain} by $R=\frac{1}{I}\frac{d I}{dn}$, it is easy to see that $R$ for this example is monotonically decreasing with $n$. As a final remark, note that this result --- for all values of $n$ --- is also consistent with the way we interpreted the quantity $e^{-I}$ in Sec.~\ref{Sec:Info-bayes}: $e^{-I}$ is equal to the standard deviation of the posterior relative to that of the prior. 
\\\\
The second proposition is the statement that the amount of information one can in principle gain is fixed regardless of the method employed to incorporate the data. We dub this statement the {\it information semi-extensivity}.
\begin{quotation}
	{\bf Proposition 2.} (information semi-extensivity):
	\\ Learning from two pieces of data does not change whether they are used all at once or one at a time, in successive steps. In equations,
	\ba 
	\label{successive-update}
	I(\{\dat_1,\dat_2\},\theo)= I(\dat_1,\theo)+I(\dat_2,\theo|\dat_1).   
	\ea 
\end{quotation}
This seemingly obvious statement reflects an information-theoretic understanding of the crucial concept in the Bayesian framework, which asserts that there is no distinction in updating probabilities (i.e., the posterior probability density remains unchanged) when using two datasets separately or together. 
\\\\
The third proposition deals with marginalization which is another  common practice in Bayesian analysis. Consider a multi-dimensional parameter space and decompose it into two disjoint subsets: $\theta=\{\theta_1 , \theta_2\}$ and suppose that only the subset $\theta_1$ is of interest. In this case, one would marginalize over $\theta_2$ (which is called the nuisance or marginalized-out subset) to obtain the marginalized likelihood given the remaining subset $\theta_1$ (which is called the marginal  subset). From the information-theoretic perspective, marginalization means that some information that is in principle accessible,  is not of interest for discriminating theories and will be lost in the marginalization procedure. This intuition may be quantified by switching the role of data and theories in  Eq.~\eqref{successive-update} and rearranging different terms, leading to the following  proposition:
\begin{quotation}
	{\bf Proposition 3.} (Marginalization): 
	\\
	By marginalizing, one loses the information that could have been gained about the nuisance parameters, given the marginal parameters. That is 
	\ba 
	\label{marginalization}
I(\theo_1,\dat)=	I(\{\theo_1,\theo_2 \},\dat) - I(\theo_2,\dat|\theo_1)\, ,
	\ea 
	where $\theo_1$ and $\theo_2$ are the marginal and the nuisance variables, respectively.\footnote{In Ref.~\cite{Bhola}, Instead of marginalization and calculating $I(\theo_1,\dat)$, it is proposed to use the conditional MI $I(\theo_1,\dat|\theo_2)$. However, we do not find a justification for this proposal. Marginalization is the standard method in this situation and it would be preferable to remain within the standard Bayesian framework.}
\end{quotation}

This proposition may be applied to the case of model selection for which a marginalization over the parameters of the models is performed (see Eq.~\eqref{model_marginalize}). In this case, Eq.~\eqref{marginalization} yields 
\ba 
I(\dat,\hat M) = I(\dat,\{\theo,\hat M \}) - I(\dat, \theo|\hat M) \, , 
\ea 
%where the conditional mutual information may be expressed by
%\ba 
%I(\dat, \theo|\hat M)  = \sum_i P(M_i) \, I(\dat, \theo|M_i)\, ,
%\ea 
%where, according to Eq.~\eqref{MI-cond-realization}), 
%\ba 
%I(\dat, \theo|M_i) = \int p(\mu,\theta|M_i) \, \ln\left(\dfrac{p(\mu,\theta|M_i)}{p(\mu|M_i) p(\theta|M_i)} \right) d \mu d\theta.
%\ea 
%Therefore, the information that is lost by being ignorant about the model parameters is the expected mutual information between  data and model parameters among all models under consideration. 
This result may be viewed in the following way: One can think of each model $M_i$ as the specification of a disjoint block in the parameter space. When model selection is of interest, one is ignorant about the details of the model (that are identified by the parameters of each model). As a result, the information about such details are lost (or, more precisely, are not taken into account). In this sense, marginalization for the purpose of model selection corresponds to a {\it coarse graining} in the parameter space such that the details of each disjoint block is removed from the analysis. 
\\\\
The last proposition that we state in this paper makes explicit the intuition that the information gain has to be invariant under the change of variables:
\begin{quotation}
	{\bf Proposition 4.} (Reparametrization): 
	\\
The information obtained about theories from an experiment is fixed, regardless of how the theories are parametrized.
 That is
	\ba 
	\label{reparam}
	I\big(f(\theo),\dat \big) = I(\theo, \dat)
	\, , \quad 
	I\big (f(\theo_1),\dat \big |g(\theo_2) \big) = I(\theo_1, \dat|\theo_2)
	\,  ,
	\ea
	for arbitrary smooth and invertible functions $f$ and $g$.
%	\footnote{The proof of this proposition simply follows from converting all conditional probabilities to joint probabilities in the definitions of MI and CMI and then using the laws of probability conservation  and  change of variables. In the multidimensional parameter space define $\tilde \theta = f(\theta)$. We have
%		\ba 
%		p(\tilde \theta) = |\det(J)|^{-1} \, p(\theta) \, , \quad d\tilde \theta = |\det (J)| \, d \theta .
%		\ea 
%		where $J$ is the Jacobian matrix ($J_{ij} = \partial \tilde \theta_i/\partial \theta_j$). Similar expressions hold for the joint probability  densities. }
\end{quotation}
This is a plausible property of  MI that might also be  practically useful for the simplification of the MI estimation. However, we stress that a natural choice of prior in one parametrization may not seem natural in others.

It is worth mentioning that the invariance of conditional MI, according to Eq.~\eqref{reparam}, does not necessarily hold under more general reparametrization such as $\{\theo_1 ,\theo_2\} \to h(\{\theo_1 ,\theo_2\})$ for some invertible function $h$. This has to be the case since such a general reparametrization may mix the variables for which MI is of interest and the ones  which MI is conditioned on. Note also that a similar reparametrization invariance holds for the other random variable $\dat$ but it is of less interest in this paper since this parameter corresponds to the actual measurement.

\section{Conclusion and outlook}
\label{Sec:conclusion}
In this paper, we proposed that mutual information (MI) is a suitable quantity to study in the context of Bayesian forecasting for a future experiment. We  argued that it can be used in both situations of interest, i.e., for parameter estimation and model selection. For parameter estimation, we  argued that MI quantifies the expected reduction in the dispersion of the posterior  probability density relative to that of the prior probability density. For model selection, we showed that MI quantifies the expected fraction of potentially available information that may be extracted by a future experiment. 

One possible objection to the method proposed in this paper is that it relies on the choice of prior. However, we note that this is not unique to this method and is inherent in any Bayesian inference. While the dependence on the prior may also have some advantages, we assert that --- as argued in Ref.~\cite{Guth:2023hbx} --- the frequentist inference is not completely free from the prior choice either, although this dependence is more concealed.

Note that the proposed method being independent of the choice of  fiducial parameters --- which we viewed as an advantage  --- may underestimate the constraining capability of the real experiment, when a true theory is indeed realized. However, this is an obvious consequence of the lack of prior knowledge about the true theory. If one has a higher {\it a priori} credence on some theory than the others, it is directly reflected in the prior probabilities which then allows that theory to contribute more to MI (see Eq.~\eqref{MI-conditional}). On the other hand, if one is interested in forecasting for a future experiment given a true theory, the information-theoretic framework  may still be adequate. As a natural proposal, one may  consider the expected change of uncertainty in the parameter space conditioned on the true theory. That is, denoting the true theory by $\theta_0$, in analogy with Eq.~\eqref{MI_DeltaH}, one may define
\ba 
I_{\theta_0} (\dat,\theo) \equiv  \int \Delta H(\theo|\mu) \, p(\mu|\theta_0) d\mu.
\ea 
Note that this quantity differs from the MI between two random variables conditioned on a specific realization of another random variable defined  in Eq.~\eqref{MI-cond-realization}. We will not pursue  this proposal further in this paper. However, we share the interesting observation that in the two concrete examples presented in this paper that led to Eqs.~\eqref{MI-posterior-width} and \eqref{MI-n-measurements} for MI, the above quantity is independent of $\theta_0$ and coincides precisely with the MI for those examples. This proposal deserves further study which we leave for future works.

Another question that arises concerning the proposed method is what values of MI leads to the conclusion that conducting an experiment would be valuable.  We used base-$e$ logarithms to define MI. Thus, $I=\ln 2\simeq 0.69$ (in {\it nats}) corresponds to one bit of data. One may decide that $I \lesssim \ln 2$  suggests that the experiment would not be informative and one would require $I \gtrsim \ln 2$ and ideally $I \gg \ln 2$. This is not necessarily a sharp and objective boundary but rough ideas may be obtained about the meaning of $I=\ln 2$ by analyzing a simple example. Consider the following one dimensional Gaussian prior and  Gaussian likelihood densities: $p(\theta) = \calN_\theta(0,
\sigma_{\text{pr}}^2)$ and $p(\mu|\theta) = \calN_\mu(\theta, \sigma_L^2)$. In this case, according to Eq.~\eqref{MI-posterior-width}, if $\sigma_{\text{pr}}=\sqrt{3}\sigma_L$, we have $I=\ln 2$ which corresponds to the width of the posterior probability density being one-half of that of the prior probability density. In other words, if $I \geq \ln 2$, a theory that is {\it a priori} $1\sigma$ away from the most favored theory is expected to become at least $2\sigma$ away from it {\it a posteriori}. This may be considered as a significant improvement in the belief on the theories towards the truth and one may decide that $I \gtrsim  \ln 2$ (and not necessarily $I\gg \ln 2$) makes the experiment already worthwhile. A more conservative  boundary like $I \simeq \ln 5$ nats (or $\ln5/\ln 2\simeq 2.3$ bits) can also be considered that enables one to update a $1\sigma$ significance to $5\sigma$ significance. A more general analysis, which we leave for the future, can lead to different categories for  the ``significance of evidence", similar to what has been done e.g., for the Bayes factor, known as {\it Jeffreys' scale} \cite{jeffreys1998}. 

For model selection, the decision might be more straightforward
since the proposed method, for this case, measures the fraction of information that is expected to become available about the models (through the ratio $I/H$). A simple example may give  rough ideas about the meaning of the numbers one may encounter. Consider two models $M_1$ and $M_2$  with likelihood densities $p(\mu|M_1)=\calN_{\mu}(0,\sigma_L^2)$ and $p(\mu|M_2)=\calN_{\mu}(5\sigma_L,\sigma_L^2)$ for an arbitrary standard deviation $\sigma_L$. That is, the predicted measured value according to one theory is $5\sigma$ away from the predicted value according to the other, indicating that the two models are expected to be distinguishable by a single measurement. Assuming equal prior probabilities (leading to $H=\ln 2$), a simple numerical integration results in $\frac IH\simeq 0.975$. Therefore, in this situation, it is expected that 97.5\% of the available information to be gained by a  measurement.\footnote{These Bayesian numbers may be compared with frequentist ones in the following way: Consider $M_1$ and $M_2$ as the true model and the null hypothesis, respectively. That is, assume that the measurement is correctly predicted by $M_1$ while $M_2$ is interested to be tested by the measurement. Denoting the $p$-value for $M_2$ by $p_{\text{val}}(\mu,M_2)$, one can compute the expected $p$-value according to the true model, i.e., $\langle p_{\text{val}}(M_2) \rangle \equiv \int p_{\text{val}}(\mu,M_2) \, p(\mu|M_1) \, d\mu$. For the current example, we have $\langle p_{\text{val}}(M_2) \rangle \simeq 4\times 10^{-4}$. That is, by a measurement, the null hypothesis is expected to be rejected by about $3.5\sigma$ level of significance.} 
%This result also supports the adequacy threshold $I\simeq \ln 2$, as suggested above. 
A  detailed study of possible interpretation of the numerical values of MI would be valuable which we leave for future works.

Another application of MI which is not influenced by the above (rather arbitrary) choice of boundaries is to study how MI varies with experiment's specifications to achieve the most optimal design. This idea will be pursued in Ref.~\cite{JWST} in a cosmological situation to determine the best range of redshift for observation, if one is interested in discovering signals due to the primordial effects (such as the ones predicted by particular models of inflation).
	
On the practical side, since MI requires integration over all possible theories and all possible realizations of data, it is clear that estimation of  MI is more challenging than e.g., the analysis based on the Fisher information matrix. Further complications associated with other properties of MI also exist \cite{pmlr-v108-mcallester20a,e13040805}. On the other hand, recent technological and methodological developments --- in particular,  the machine-learning-based techniques --- make the feasibility of the estimation of MI for Bayesian forecasting a promising  future \cite{2004PhRvE..69f6138K,Jeffrey:2023stk}.

\acknowledgments
I thank Y. Akrami, M. Ansarifard, A. H. Guth, A. Mollabashi, B. Nikbakht and A. Shafieloo for useful comments and discussions. 

% \appendix
 
% \section{Proof of propositions}
% \label{app:proofs}
%
%{\bf{Proposition` 1.}} 
%
%{\bf{Proposition 2.}} 
%
%{\bf{Proposition 3.}} 
%The proof simply follows from converting all conditional probabilities to joint probabilities and then using the laws of probability conservation  and  change of variables. In the multidimensional parameter space define $\tilde \theta = f(\theta)$. We have
%\ba 
%p(\tilde \theta) = |\det(J)|^{-1} \, p(\theta) \, , \quad d\tilde \theta = |\det (J)| \, d \theta .
%\ea 
%where $J$ is the Jacobian matrix ($J_{ij} = \partial \tilde \theta_i/\partial \theta_j$). Similar expressions hold for the joint probability densities. 

%\bibliography{biblio.bib}

\end{document}